\documentclass[fleqn,usenatbib]{mnras}

\usepackage[T1]{fontenc}
\usepackage{ae,aecompl}
\usepackage{caption}
\usepackage[figuresright]{rotating}
\usepackage{graphicx}	% Including figure files
\usepackage{amsmath}	% Advanced maths commands
\usepackage{amssymb}	% Extra maths symbols

\usepackage{newtxtext,newtxmath}

\newcommand{\Ei}{$E_{\rm i}$}
\newcommand{\eps}[1]{\log\varepsilon_{\rm #1}}
\newcommand{\abun}{$\log(N_{\rm el}/N_{\rm tot})$}

\def\ione{\,{\sc i}}
\def\ii{\,{\sc ii}}

\def\vmicro{$\xi_{\rm t}$}

\newcommand{\te}{$T_{\rm eff}$}
\newcommand{\logg}{$\log{g}$}

\def\vmacro{$\zeta_{\rm RT}$}

\def\CT{$T_{\rm c}$}
\newcommand{\SME}{\textsc{sme}}

\title[Wide pairs with and without planets]
{Detailed abundances of the wide pairs of stars with and without planets: the binary systems 16~Cyg and HD~219542}

\author[T. Ryabchikova et al.]{
T.~Ryabchikova,$^1$\thanks{E-mail: ryabchik@inasan.ru} Yu.~Pakhomov,$^{1}$ L.~Mashonkina,$^{1}$ T.~Sitnova$^{1}$ \\\\ 
$^{1}$Institute of Astronomy of the Russian Academy of Sciences, Pyatnitskaya str. 48, 119017, Moscow, Russia \\
}

\date{Accepted XXX. Received YYY; in original form ZZZ}
\pubyear{2022}

\begin{document}
\label{firstpage}
\pagerange{\pageref{firstpage}--\pageref{lastpage}}
\maketitle

% Abstract of the paper
\begin{abstract}
 
We present results of the comparative analysis of the two wide binary systems -- 16~Cyg, with a giant gas planet orbiting around 16~Cyg~B, and HD~219542 without planet detected. Atmospheric parameters of the binary components and the Sun were determined using their high-resolution spectra and the \SME\ tools for automatic spectral analysis. By applying the synthetic spectrum method, we derived abundances of 29 and 23 chemical elements in 16~Cyg and HD~219542, respectively. For 19 of these elements, our results are based on the non-local thermodynamic equilibrium (NLTE) line formation. For both 16~Cyg and HD~219542, we obtained a small abundance difference between the A and B components: +0.019$\pm$0.012 and $-0.014\pm$0.019, respectively, suggesting only a weak influence of the giant gas planet formation on chemical composition of the host star atmosphere. For HD~219542~A and B, trends of the relative-to-solar abundances with the dust condensation temperature are similar to the literature data for the solar analogues without detected planets. The components of 16~Cyg reveal very similar behaviour of [X/H] with the condensation temperature, however, it is different from that for HD~219542. This indicates a specific chemical composition of the cloud from which the 16~Cyg binary system formed.
 
%The abstract should briefly describe the aims, methods, and main results of the paper.
%It should be a single paragraph not more than 250 words (200 words for Letters).
%No references should appear in the abstract.
\end{abstract}

\begin{keywords}
(stars:) binaries: visual -- stars: atmospheres -- stars: abundances -- star: evolution
\end{keywords}

%%%%%%%%%%%%%%%%%%%%%%%%%%%%%%%%%%%%%%%%%%%%%%%%%%

%%%%%%%%%%%%%%%%% BODY OF PAPER %%%%%%%%%%%%%%%%%%

\section{Introduction}

With the development of modern spectroscopic instruments precise abundance studies of the  pairs of stars with similar characteristics (spectral class, proper motion, kinematics) including  
wide binary systems (hereafter wide pairs) as well as comoving pairs became a very important task. 
The stars in such systems are expected to have a common origin, hence, a common chemical composition. Therefore even small observed abundance differences provide a difficulty for the theory and request a proper explanation. One of the reasons for the chemical inhomogeneity may be an incomplete mixing of the interstellar medium. At the same time, the discovery of the chemically inhomogeneous binary systems where one of the stars is a planet host, allows us to attribute an overall abundance difference between the components to a chemical signature of planet formation. 

High precision differential abundance analyses of the stars in wide bound pairs have a potential to discover 'fingerprints' of the confirmed or possible planets and planet formation  

In past years, such studies were performed for both individual pairs and groups of wide pairs with/without detected planets.     

\citet{2001AA...377..123G} analysed six wide pairs (separations of few hundreds AU) without detected planets. No abundance difference between the components was found in the four systems, while a significant abundance difference was derived for HD~219542 and HD~200466.  This kind of strictly differential abundance analysis was extended to larger groups of wide binaries with the similar solar-like components\footnote{We assume here the stars with the effective temperatures $\pm$500-700~K, surface gravity $\pm$0.4~dex and  metallicity $\pm$0.3~dex from the solar parameters.}
\citep{2004AA...420..683D,2006AA...454..581D, 2020ApJ...888L...9N,2020MNRAS.492.1164H}. 
\citet{2019ApJ...871...42A} and \citet{2021ApJ...921..118N} tested the abundance consistency for 31 wide binaries and 31 comoving pairs. In both studies, the similar chemistry of the components, at the level of 0.1~dex or even less, was found. However, a differential abundance analysis of the comoving pairs HD~240429/HD~240430 \citep{2018ApJ...854..138O} and HIP~34407/HIP~34426 \citep{2019MNRAS.490.2448R} reveals an enhancement of the refractory elements (the dust condensation temperature \CT\ $>$ 1200~K), at the level of up to 0.2~dex. For HD~240429/HD~240430, it was interpreted as an accretion of rocky material by one of the components after the birth.   
The largest collection consisting of 107 objects was recently analysed by 
\citet{2021NatAs...5.1163S}. The authors conclude that about a quarter of the solar-like objects show an excess of the refractory elements in one of the components that may be caused by a planet engulfment event.

\citet{2021MNRAS.508.1227L} considered some additional effects that can produce the observed relative abundance difference in binary systems. First, an incomplete mixing of the interstellar medium can result in slightly different chemical composition of the binary components. Next is atomic diffusion that alters the surface abundances differently for different elements in the case of fairly large difference in atmospheric parameters of the components. 

Different research groups performing the differential abundance analyses for common wide pairs report sometimes the controversial results. 
For example, multiple studies were performed for the 16~Cyg system --  one of the best studied binaries with detected planets.
\citet{2011ApJ...737L..32S} did not find any abundance difference between the two components, while \citet{2011ApJ...740...76R} derived a difference of +0.04~dex between A and B, which does not depend on the condensation temperature (\CT). Based on the better quality observations obtained with two different spectrographs, \citet{2014ApJ...790L..25T} and \citet{2019AA...628A.126M} 
confirmed the result of \citet{2011ApJ...740...76R} that 16~Cyg~A is more metal rich than 16~Cyg~B. However, \citet{2014ApJ...790L..25T} and \citet{2019AA...628A.126M} got different trends of the relative abundances (A-B) with \CT\ suggesting a different interpretation of the planet formation effects. 

Another example is a wide pair HD~219542 without detected planets. A significant abundance difference of $\sim$0.09~dex between the components was reported by \citet{2001AA...377..123G} and confirmed by \citet{2003PASJ...55.1005S} who used different spectral observations. However, using the same spectra as \citet{2001AA...377..123G}, \citet{2004AA...420..683D} did not find any abundance difference between the components. With another set of spectra, \citet{2006AA...454..581D} confirmed an absence of the metallicity difference.

In all the cited spectroscopic studies, atmospheric parameters of the binary components were derived using the Fe\ione\ excitation equilibrium and Fe\ione/Fe\ii\ ionisation equilibrium methods and the equivalent width measurements. Our practice shows that, in the solar-like and cooler stars of enhanced metallicity, it is very difficult and sometimes impossible to find a reasonable number of the Fe\ione\ and Fe\ii\ lines, which are totally free from blending. In the red spectral regions where the atomic line blending is less severe, numerous weak molecular lines start to play an important role. 

This study deals with the 16~Cyg and HD~219542 binary systems and aims to perform a detailed abundance analysis of their components. The giant gas planet around 16~Cyg~B was discovered by \citet{1997ApJ...483..457C}. The A component of 16~Cyg is itself a close binary, with the M dwarf companion \citep{2006ApJ...646..523R}, but no planet detected. No planet was discovered in the HD~219542 binary system. We use the high-quality stellar spectra in a wide spectral region from 4400 to 9000~\AA, determine stellar atmosphere parameters with the automatic spectral analysis tools based on a variety of spectroscopic indicators of effective temperature and surface gravity, and, for the first time, derive elemental abundances of 16~Cyg and HD~219542 by the synthetic spectrum method with taking into account the departures from the local thermodynamical equilibrium -- LTE (the NLTE approach).

The paper is organized as follows. Observations are described in Section~\ref{sect:obs} and determinations of stellar atmosphere parameters in Section~\ref{sect:param}. Abundances of 29 elements in the Sun and the components of the 16~Cyg and HD~219542 binary systems are derived in Section~\ref{sect:abund}. The abundance differences between the stars are discussed in Section~\ref{sect:Discussion}. Our conclusions are summarised in Section~\ref{sect:Conclusions}.

\section{Observations}\label{sect:obs}

\underline{16~Cyg}. Two sets of the high-resolution spectra of the 16~Cyg system were downloaded from the archives. The first one is Canadian Astronomy Data Centre (CADC) archive \citep{1994ASPC...61..123C}. The 16~Cyg~A and B stars were observed on 2013-06-29 in the frames of the Program 13AB06 with the Echelle SpectroPolarimetric Device for Observation of Stars (ESpAdOnS)  spectrograph mounted at the 3.6~m Canada-France-Hawaii Telescope  \citep[CHFT,][]{2003ASPC..307...41D}. Spectral resolving power is R = 81\,000, and wavelength coverage is 3696 -- 10482~\AA. Total exposure times were 840~s and 1050~s for the A and B components, respectively. Signal-to-noise ratio (S/N) varies from 200 to 500 for different spectral ranges. In addition, we downloaded the spectrum of the solar light reflected from the asteroid Vesta that was observed in the frames of the same observational program. More details on the ESPaDOnS/CFHT spectra of 16~Cyg and Vesta can be found in \citet{2014ApJ...790L..25T}. 

%y were obtained with two different spectrographs: ) at Mauna Kea and  
Spectra of 16~Cyg observed with the High Dispersion Spectrograph \citep[HDS,][]{2002PASJ...54..855N} mounted on the 8.2~m Subaru Telescope of the National Astronomical Observatory of Japan were downloaded from the Subaru Mitaka Okayama Kiso Archive system \citep[SMOKA,][]{2002ASPC..281..298B}. R = 160\,000 in the visual region. More details on these spectra can be found in \citet{2019AA...628A.126M}. 

\underline{HD~219542.} Spectra of both components were downloaded from the SMOKA archive. They were observed with HDS/Subaru on 2002-12-18 in the frames of the Proposal ID o02201. R = 89\,000.
Total exposure times were 900~s and 1200~s for the A and B components, respectively, resulting in S/N of about 250-350. Wavelength coverage is 4326 -- 5821~\AA\ in 35 orders for the blue region and 5695 -- 7209~\AA\ in 22 orders for the red region.  
Spectra were reduced with the help of the MIDAS program complex \citep{1992ASPC...25..115W}. 
We used the same spectral data as in \citet{2003PASJ...55.1005S}. 

All the downloaded spectra were normalised to the continuum level by fitting a global smooth function to the observed spectrum using a procedure similar to that described in the Appendix~A of \citet{2018A&A...613A..60R}. Everywhere except H-lines we used 3d order polynomial approximation for a smooth function.

Comparison between the ESpAdOnS/CFHT and HDS/Subaru spectra of 16~Cyg showed that the latter spectra have a number of gaps between the orders and low S/N in their red part. Therefore we used only the ESpAdOnS spectra in our analysis. Moreover, the resolving power of the ESpAdOnS/CFHT spectra for 16~Cyg is similar to that for the HDS/Subaru spectra of HD~219542. This is favourable for a comparison study of the two systems. 

\section{Atmospheric parameters}\label{sect:param}

\begin{table*}
	\caption{Atmospheric parameters of the programme stars with the errors estimated by two methods.}\label{param}
	\begin{tabular}{r ccccc cccc cccc cccc ccc}
		\hline
		Star&\te&$\sigma_1$&$\sigma_2$&&logg&$\sigma_1$&$\sigma_2$&&$[M/H]$&$\sigma_1$&$\sigma_2$&&\vmicro&$\sigma_1$&$\sigma_2$&&\vmacro&$\sigma_1$&$\sigma_2$\\
		\hline
		Sun (Vesta) & 5778 & 70 & 19 &&  4.44  & 0.16 & 0.02 && 0.003 & 0.053 & 0.012 &&  0.86  & 0.24 & 0.03 && 3.59  & 0.81 & 0.61 \\
		16 Cyg A    & 5829 & 42 & 30 &&  4.33  & 0.13 & 0.04 && 0.110 & 0.038 & 0.023 &&  0.99  & 0.15 & 0.05 && 4.21  & 0.53 & 0.39 \\
		16 Cyg B    & 5760 & 40 & 34 &&  4.39  & 0.11 & 0.04 && 0.074 & 0.033 & 0.024 &&  0.90  & 0.14 & 0.06 && 3.32  & 0.54 & 0.39 \\ 
		HD 219542 A & 5880 & 59 & 32 &&  4.39  & 0.19 & 0.05 && 0.104 & 0.049 & 0.027 &&  1.08  & 0.13 & 0.05 && 3.89  & 0.58 & 0.32 \\
		HD 219542 B & 5753 & 55 & 37 &&  4.47  & 0.16 & 0.05 && 0.134 & 0.049 & 0.030 &&  1.01  & 0.13 & 0.06 && 3.41  & 0.56 & 0.57 \\
		\hline
	\end{tabular}
\end{table*}

For each star, its atmospheric parameters: effective temperature (\te), surface gravity (\logg), metallicity ([M/H]), micro- (\vmicro) and macroturbulent (\vmacro) velocities were obtained with the Spectroscopy Made Easy (\SME) IDL program package designed for the automatic spectral analysis \citep{1996AAS..118..595V,2017A&A...597A..16P}. \SME\ was successfully applied to determinations of the atmospheric parameters and abundances of different stars \citep{2005ApJS..159..141V, 2016ApJS..225...32B}. The accuracy of the atmospheric parameter determinations with \SME\ was investigated by \citet{2016MNRAS.456.1221R} who showed that effective temperatures, surface gravities, and metallicities agree within $\pm65$~K, $\pm0.12$~dex, and $\pm0.04$~dex with other spectroscopic determinations. The \SME\ analysis is based on fitting the synthetic spectra to the observations in chosen spectral regions by varying atmospheric parameters. 
Simultaneous optimization of free parameters is performed with a $\chi^2$ minimization algorithm described by  
\citet{Marquardt_1963} and \citet{Press_1986}.
This procedure requires a careful choice of spectral lines (line mask) with a different sensitivity to the variations in atmospheric parameters. 

\begin{table*}
	\caption{Published atmospheric parameters of the programme binary systems. }
	\label{param_published}
	\tabcolsep=3.0pt
	\begin{tabular}{ccccccl}
		\hline
		\multicolumn{7}{c}{16~Cyg}\\
		\multicolumn{2}{c}{\te} & \multicolumn{2}{c}{\logg} &\multicolumn{2}{c}{[Fe/H]}& Reference\\
		A& B& A& B& A& B &\\
		\hline
		5795(20)  & 5760(20)   & 4.30(06)~~ &4.40(06)~~ &0.04(02)~~~&0.06(02)~~~ & \citet{2000AJ....119.2437D}\\
		5745(40)  & 5685(40)   & 4.21(07)~~ &4.26(08)~~ &0.10(03)~~~&0.07(03)~~~ & \citet{2001ApJ...553..405L}\\
		5825(50)  & 5750(50)   & 4.33(07)~~ &4.34(07)~~~&0.096(026) &0.052(021)  & \citet{2009AA...508L..17R}\\ 
		5813(18)  & 5749(17)   &4.282(017)  &4.328(017) &0.103(023) &0.069(026)  & \citet{2011ApJ...740...76R}\\
		5796(34)  & 5753(30)   &4.38(12)~~  &4.40(12)~~~&0.07(05)~~~&0.05(05)~~~ & \citet{2011ApJ...737L..32S}\\
		5830(7)~  & 5751(6)~   &4.30(02)~~  &4.35(02)~~~&0.101(008) &0.054(008)  & \citet{2014ApJ...790L..25T}\\
		5816(10)  & 5763(10)   &4.291(010)  &4.356(010) &0.093(007) &0.062(007)  & \citet{2017AA...608A.112N}\\
		5832(5)~  & 5763(5)~   &4.310(014)  &4.360(014) &0.103(004) &0.063(004)  & \citet{2019AA...628A.126M} \\
		5839(42)  & 5809(39)   &            &           &           &            & \citet{2013MNRAS.433.1262W}\\
		          &            &4.292(002)  &4.359(002) &           &            & \citet{2015ApJ...811L..37M}\\
		5864(48)  & 5814(59)   &4.302(014)  &4.373(015) &0.15(05)~~ &0.12(03)~~  & \citet{2022AA...658A..47K}\\
		\hline
		\multicolumn{7}{c}{HD~219542}\\
		\multicolumn{2}{c}{\te} & \multicolumn{2}{c}{\logg} &\multicolumn{2}{c}{[Fe/H]}& Reference\\
		A& B& A& B& A& B &\\
		\hline
		5989(50)  &5713(50)    &4.37        &4.38      &0.293(014) &0.201(013)  & \citet{2001AA...377..123G}\\
		5830(30)  &5600(30)    &4.45(10)    &4.40(10)  &0.13(04)~~ &0.08(04)~~  & \citet{2003PASJ...55.1005S}\\
		5859(30)  &5691(30)    &4.34(10)    &4.42(10)  &0.14(04)~~ &0.14(04)~~  & \citet{2004AA...420..683D}\\
		5826(30)  &5674(30)    &4.32(10)    &4.40(10)  &0.06(04)~~ &0.07(04)~~  & \citet{2006AA...454..581D}\\
		\hline
	\end{tabular}

{\it Note.} Numbers in parenthesis indicate errors in the last digits.
\end{table*}

The same six spectral regions, that is 4485--4590~\AA, 4820--4880~\AA\ (H$\beta$), 5100--5200~\AA, 5600--5700~\AA, 6100--6200~\AA, and 6520--6580~\AA\ (H$\alpha$), and the same line masks as in \citet{2016MNRAS.456.1221R} were used in the fitting procedure. 
These regions contain both atomic and molecular lines sensitive to effective temperature variations in solar-like stars (H$\beta$ and H$\alpha$ lines, Swan system of molecular C$_2$ lines, V\ione, Ti\ione, Co\ione\ lines) as well as the lines sensitive to gravity variations (numerous Ti\ione-Ti\ii\ and Fe\ione-Fe\ii\ lines with accurate laboratory line parameters, the Mg\ione\ triplet 5167-83 \AA).
For HD~219542, we implemented two additional regions: 5520-5535~\AA\ and 5710-5716~\AA\ (Mg\ione). These regions cover the neutral magnesium lines which help in refining the surface gravity being combined with  sensitive to the pressure effects. Atomic and molecular line parameters for spectral synthesis were extracted from a recent version of the Vienna Atomic Line Database ({\sc VALD3})\footnote{http://vald.inasan.ru} which provides isotopic and hyperfine structure components of the spectral lines for a number of elements \citep{2019ARep...63.1010P}. The data source for C$_2$ lines is \citet{BBSB}. 

The \SME\ package employs three grids of model atmospheres: {\sc ATLAS9} \citep{2003IAUS..210P.A20C}, {\sc LLmodels} \citep{2004AA...428..993S}, and {\sc MARCS} \citep{2008AA...486..951G}. All these models assume plane-parallel (one-dimensional, 1D) geometry and the local thermodynamical equilibrium (LTE). Our calculations were done with the {\sc LLmodels} models because they have a finer grid of the depth points over the atmosphere.

The derived atmospheric parameters are listed in Table~\ref{param} together with the error estimates calculated by two methods.  
The first one ($\sigma_1$) is described in papers by \citet{2016MNRAS.456.1221R} and \citet{2017A&A...597A..16P}. It is based on analysis of the cumulative distribution for each free parameter constructed using all pixels selected for fitting by the mask. For each pixel the partial derivative of the synthetic spectrum with respect to the free parameter is computed and the parameter change needed to cancel the difference between synthetic and observed spectrum is estimated.  The distribution function usually has very wide wings due to insensitivity of some pixels to the selected parameter variations or due to erroneous/missing atomic and molecular data.
However, the central part of the corresponding probability distributions is not too far from a Gaussians \citep[see Fig.~1 and Fig.~2 in][]{2016MNRAS.456.1221R} and, hence, can be used for estimating realistic uncertainties of free parameters. This method assumes that the model errors rather than the measurement uncertainties dominate the resulting values of free parameters. Model errors include data reduction glitches, uncertainties in line data, neglected NLTE effects, errors in assumed instrumental profile, continuum normalisation, etc.
The second set of errors ($\sigma_2$) is more of a performance assessment of the optimisation algorithm implemented in \SME. It would give realistic uncertainties of the free parameters provided that our model of stellar spectrum is 'perfect' in a sense that it will converge to the observations in every wavelength point with increasing S/N of the data. In reality, we have to deal with errors in atomic and molecular data, limitations of atmospheric models, instrumental defects, continuum normalisation, and other systematic issues, thus the level of reduced $\chi^2$ never reach 1 and the uncertainties derived from the covariance matrix are highly underestimated even though they still show the level of convergence. The uncertainties based on cumulative distribution analysis are overestimated since we treat every free parameter independently assuming its responsibility for all residuals of spectral fitting. Obviously, for parameters that affect less data points (e.g. abundance of a particular species) our estimate will be highly exaggerated. The true uncertainties in free parameters are somewhere in between the two estimations.

Reliability of the obtained atmospheric parameters is confirmed by a perfect agreement of \te\ = 5778~K and \logg\ =4.44, which were derived using the spectrum of the solar light reflected from the  asteroid Vesta, with the canonical solar values. 

Table~\ref{param_published} collects the published atmospheric parameters for both wide pairs.  
Atmospheric parameters of the 16~Cyg components derived with the \SME\ automatic package agree well with most of the spectroscopic determinations in the literature. 
An independent method of effective temperature determinations is provided by direct measurements of the star's diameter and frequency-integrated flux. 
Recently, the 16~Cyg system was included in a group of the benchmark dwarf stars for direct stellar diameter measurements by means of interferometry \citep{2022AA...658A..47K}. Similar measurements were made earlier by \citet{2013MNRAS.433.1262W}. Radii and effective temperatures of the components derived in both studies agree within the error limits. 
The effective temperatures obtained in this study for the 16~Cyg components are consistent within the error bars with those based on the interferometric measurements of \citet{2022AA...658A..47K}. 
For~HD~219542A, both \te\ and \logg\ are slightly higher, but still
agree within the errors with the corresponding parameters from the
last three publications in Table~\ref{param_published}. For HD~219542B, the \SME\ procedure
returns the higher effective temperature, by 79~K to 153~K. We note a
rather large dispersion of the literature parameters for HD 219542B
compared to that for HD 219542A.
According to our analysis, both binary systems have similar, slightly supersolar metallicities.

\section{Chemical abundances}\label{sect:abund}

\begin{table}
	\caption{Model atoms used in our NLTE calculations.} \label{model_atoms}
	% \centering
	\begin{tabular}{ll}
		\hline \noalign{\smallskip}
		Species & Reference  \\
		\hline \noalign{\smallskip}
		C\ione &   \citet{2015MNRAS.453.1619A} \\
		N\ione &   Neretina (in prep) \\
		O\ione &   \citet{2018AstL...44..411S}  \\
		Na\ione &   \citet{alexeeva_na}         \\
		Mg\ione & \citet{mash_mg13}  \\
		Al\ione & \citet{mash_al2016} \\
		Si\ione -\ii  &   \citet{2020MNRAS.493.6095M} \\
		K\ione & \citet{2020AstL...46..621N}  \\
		Ca\ione -\ii &   \citet{2017AA...605A..53M}   \\
		Ti\ione -\ii &   \citet{sitnova_ti}   \\
		Fe\ione -\ii &   \citet{mash_fe}  \\
		Zn\ione -\ii &  \citet{sitnova_zn}  \\
		Zr\ione -\ii &   \citet{Velichko2010_zr}  \\
		Ba\ii & \citet{2019AstL...45..341M}  \\
		\noalign{\smallskip}\hline \noalign{\smallskip}
	\end{tabular}
\end{table}

\subsection{Method of calculations}

\begin{table*}
\caption{NLTE abundances from individual lines in the Sun, 16~Cyg A and B components, HD~219542 A and B components together with the line atomic parameters and the references to them.}\label{line-list}
\begin{tabular}{lrrrrrrr rrr lcc}
\hline
&\multicolumn{7}{c}{\abun}  &  &  &  & \multicolumn{3}{c}{References} \\
\cline{2-8}\cline{12-14}\\
Ion  &
\multicolumn{1}{c}{Sun} &
\multicolumn{2}{c}{16 Cyg, \SME} & 
\multicolumn{2}{c}{16 Cyg, interf.} & 
\multicolumn{2}{c}{HD 219542, \SME} &
\multicolumn{1}{c}{$\lambda$, \AA} &
\multicolumn{1}{c}{\Ei, eV} &
\multicolumn{1}{c}{log\,$gf$} & 
\multicolumn{1}{c}{$gf$} & 	HFS & IS \\
&
\multicolumn{1}{c}{(Vesta)}&
\multicolumn{1}{c}{A} & 
\multicolumn{1}{c}{B} &  
\multicolumn{1}{c}{A} & 
\multicolumn{1}{c}{B} & 
\multicolumn{1}{c}{A} & 
\multicolumn{1}{c}{B} & & & & & &  \\
\hline
& \multicolumn{1}{c}{2} & \multicolumn{1}{c}{3} & \multicolumn{1}{c}{4} & \multicolumn{1}{c}{5} & \multicolumn{1}{c}{6} & \multicolumn{1}{c}{7} &	 \multicolumn{1}{c}{8} & \multicolumn{1}{c}{9} & \multicolumn{1}{c}{10} & \multicolumn{1}{c}{11} & \multicolumn{1}{c}{12} & \multicolumn{1}{c}{13} & \multicolumn{1}{c}{14} \\
\hline
Li\ione  & -10.93~ & -10.58~ & -11.29~ & -10.64~ & -11.41~ &  -9.84~ & -10.85~ & 6707.764 &  0.000 & -0.002 &       YD   & BBE OAO      & REB\\
Li\ione  &         &         &         &         &         &         &         & 6707.914 &  0.000 & -0.303 &       YD   & BBE OAO      & REB\\
C\ione   &  -3.575 &  -3.566 &  -3.556 &  -3.603 &  -3.599 &  -3.512 &  -3.553 & 5052.144 &  7.685 & -1.303 &      NIST10&              & \\
O\ione   &  -3.311 &  -3.305 &  -3.288 &  -3.312 &  -3.282 &  -3.253 &  -3.238 & 6300.304 &  0.000 & -9.720 &     FIS98  &              & \\
\hline
\end{tabular}

{\it Note.} This table is available in its entirety in a machine-readable form in the online journal. A portion is shown here for guidance regarding its form and content. 
$\log\varepsilon_{\rm el} = \log(N_{\rm el}/N_{\rm H})$ +12 = $\log(N_{\rm el}/N_{\rm tot})$ + 12.04 in the atmospheres with the solar He abundance.
YD = \citet{YD}; BBE = \citet{BBE}; OAO = \citet{OAO}; REB = \citet{REB}; NIST10 = \citet{NIST10}; FIS98 = \citet{FIS98}. Close blends are marked by colon.
\end{table*}

Abundances of 14 chemical elements (C, N, O, Na, Mg, Al, Si, K, Ca, Ti, Fe, Zn, Zr, Ba) were derived from the NLTE calculations. We applied comprehensive model atoms from our earlier studies, which are listed in Table~\ref{model_atoms}. The model atoms were build up using the most up-to-date atomic data on energy levels, transition probabilities, photoionization cross-sections, rate coefficients for inelastic collisions with electrons and neutral hydrogen atoms. For C\ione, the model atom was updated by including the rate coefficients for the C\ione\ + H\ione\ collisions from \citet{2019AA...624A.111A}. In the case of using the approximate Drawinian rates for collisions with H\ione\ \citep{Drawin1969,Steenbock1984}, we applied the scaling factor $S_{\rm H}$, which was adopted as $S_{\rm H}$ = 1 for Ti\ione-\ii, $S_{\rm H}$ = 0.5 for Fe\ione-\ii, and $S_{\rm H}$ = 0.1 for Zr\ione-\ii.
The coupled radiative transfer and statistical equilibrium equations were solved with a revised version of the \textsc{detail} code  \citep{Giddings81,Butler84}. The update was presented by \citet{mash_fe}.

Abundances from individual lines were determined by fitting the synthetic line profiles to the observed spectrum. The synthetic spectra were calculated with the \textsc{synthV\_NLTE} code \citep{Tsymbal2018} included in the \textsc{idl} routine \textsc{binmag6} \citep{2018ascl.soft05015K}. \textsc{synthV\_NLTE} implements the departure coefficients from \textsc{detail},  
allowing us to obtain the best fit to the observed line profile with the NLTE effects taken into account. 
\textsc{binmag6} uses the non-linear least-squares fitting code \citep{2009ASPC..411..251M} for free parameters optimization. The influence of the nearby spectral lines is accounted for in the minimization procedure.

For S, Cr, Mn, Co, and Rb, we determined the LTE abundances and calculated the NLTE abundances by applying the NLTE abundance corrections, $\Delta_{\rm NLTE} = \eps{NLTE} - \eps{LTE}$, from the literature. For the S\ione\ lines, we used $\Delta_{\rm NLTE}$s from \citet{2009ARep...53..651K}.
For lines of Cr\ione, Mn\ione, and Co\ione, $\Delta_{\rm NLTE}$s were taken from the NLTE\_MPIA website\footnote{\tt https://nlte.mpia.de/gui-siuAC\_secE.php} that provides the data from the NLTE calculations of \citet[][Cr\ione]{2010A&A...522A...9B}, \citet[][Mn\ione]{2019A&A...631A..80B}, and \citet[][Co\ione]{2010MNRAS.401.1334B}. As indicated at the NLTE\_MPIA website, the model atom of Co\ione\ was updated by including the rate coefficients for the Co\ione\ + H\ione\ collisions from quantum mechanical calculations. The departures from LTE can be neglected for lines of Cr\ii, following conclusions of \citet{2010A&A...522A...9B}. 
\citet{2020AstL...46..541K} provide the NLTE corrections for lines of Rb\ione.

The lithium NLTE abundances were derived based on the Li\ione\ 6707.8~\AA\ line profile fitting in \SME\ that implements the grid of the departure coefficients for Li\ione\ pre-calculated according to \citet{2009A&A...503..541L}.

Abundances of the remaining elements (Sc, V, Ni, Cu, Y, La, Ce, Nd, Sm) were determined under the LTE assumption. The NLTE effects for Sc\ii, Ni\ione, and Cu\ione\ are studied in the literature, however, we did not find the NLTE abundance corrections for the spectral lines and the stars, analysed in this study. For the solar Sc\ii\ lines, $\Delta_{\rm NLTE}$ = $-0.01$ to $-0.02$~dex for Sc\ii\ 5667, 5669, 5684~\AA\ and $\Delta_{\rm NLTE} = -0.08$~dex for Sc\ii\ 5526 and 5657~\AA, according to \citet{Zhang2008_sc}. 
Minor NLTE effects are found by \citet{cu1_nlte_shi} for the solar Cu\ione\ lines in the visible spectral range, with $\Delta_{\rm NLTE}$ = +0.01 to +0.02~dex, while $\Delta_{\rm NLTE} = -0.05$~dex for Cu\ione\ 8092~\AA. Neither \citet{2013ApJ...769..103V}, nor \citet{2021MNRAS.508.2236B} report the NLTE abundance corrections for the Ni\ione\ lines analysed in this paper. 

We note that accounting for the NLTE effects for our LTE chemical species would affect the abundance differences between the star and the Sun and between the two components of the binary systems by no more than 0.01 dex because of their close atmospheric parameters. Our test calculations for lines of O I and Ba II in the binaries under investigation showed that, compared with LTE, NLTE reduces the abundance discrepancy between the components A and B.
We present also the carbon abundances from the C$_2$ molecular lines (the Swan band at 5100-5200~\AA), which were obtained in the \SME\ procedure. Hereafter, they are denoted as C(mol).  

The list of the used atomic lines together with the line data and the references to oscillator strengths ($gf$), isotopic structure (IS) and hyperfine structure (HFS) data are given in Table~\ref{line-list}. 

\subsection{Abundance results}

NLTE abundances, $\log(A)$ = \abun, from individual lines in the Sun, A and B components of 16~Cyg, and A and B components of HD~219542 are presented in Table~\ref{line-list}. They were derived using the atmospheric parameters determined in this study (Table~\ref{param}). 

For 16~Cyg~A and B, we also employed \te /\logg\ derived by \citet{2022AA...658A..47K} from non-spectroscopic methods, namely, effective temperatures based on their own interferometric measurements of stellar diameters and surface gravities based on stellar masses from the Dartmouth stellar evolutionary tracks \citep{2008ApJS..178...89D}. 
We note that the adopted \logg\ values agree within 0.01~dex with the asteroseismological measurements of \citet{2015ApJ...811L..37M} given in Table~\ref{param_published}. 
%Metcalfe et al. (2015, see Table 2). 
Hereafter, we refer to these non-spectroscopic atmospheric parameters as interferometric ones, although this is true for the effective temperatures only.

The average elemental abundances, their statistical errors ($\sigma$), and the number of used lines (N) are presented in Table~\ref{Vesta-16Cyg} (Sun, 16~Cyg~A, 16~Cyg~B) and Table~\ref{HD219542-abun} (HD~219542~A, HD~219542~B). Here, $\sigma$ is the dispersion in the single line measurements about the mean. 
In order to calculate the mean relative-to-solar abundances, [X/H], we applied a line-by-line differential approach, in the sense that stellar line ]abundances were compared with individual abundances of their solar counterparts. The abundance differences [A-B] = [X/H]$_{\rm A}$ -- [X/H]$_{\rm B}$ are also calculated with a line-by-line differential approach.
The last column of Table~\ref{Vesta-16Cyg} indicates the dust condensation temperatures \CT\ for 50\%\ trace element condensation, as calculated by \citet{2003ApJ...591.1220L} for a solar-system chemical composition gas.
Actually, metallicity of the program stars is slightly higher than the solar one, by about 0.1~dex. This can lead to a change in \CT\ by no more than 30--50~K, according to the simulations by \citet{2010ApJ...715.1050B} for extrasolar planetary systems of supersolar metallicity. 

The obtained solar abundances are fairly consistent with the most recent solar photosphere \citep{2021A&A...653A.141A} and meteoritic \citep{2021SSRv..217...44L} abundances. For 28 elements, in common, the average abundance differences (this study -- literature) amount to 0.022$\pm$0.053~dex and 0.027$\pm$0.063~dex, respectively.
This is well within the uncertainty of the best solar abundance determinations. 
The solar abundance comparisons give a credit to our stellar abundance results.

%\begin{landscape}
\begin{sidewaystable*}
	\centering
%	\begin{table}
%	\captionsetup{width=3in}
		\caption{Average NLTE abundances of the Sun (Vesta), 16~Cyg~A, and 16~Cyg~B.}\label{Vesta-16Cyg}
		%			Errors correspond to the 1$\sigma$ scatter of the line-by-line abundances.} 
	\begin{small}
		\tabcolsep=2.0pt
		\renewcommand{\arraystretch}{1.2}
		\begin{tabular}{l rrc @{\;\;\;\;} rrccc @{\;\;\;} rrccc @{\;\;~\;} cc @{\;\;\;~\;} rrccc @{\;\;\;} rrccc @{\;\;~\;} cc @{\;\;\;} c}
			\hline
			& \multicolumn{3}{c}{ Sun (Vesta)}  &  \multicolumn{5}{c}{16 Cyg~A}          & \multicolumn{5}{c}{16 Cyg~B}& \multicolumn{2}{c}{16 Cyg}& \multicolumn{5}{c}{16 Cyg~A}& \multicolumn{5}{c}{16 Cyg~B}& \multicolumn{2}{c}{16 Cyg} & Tc\\
			& \multicolumn{3}{c}{SME}  &  \multicolumn{5}{c}{SME}          & \multicolumn{5}{c}{SME}& \multicolumn{2}{c}{SME}& \multicolumn{5}{c}{Interferometry}& \multicolumn{5}{c}{Interferometry}& \multicolumn{2}{c}{Interferometry} & \\
			Elem    & N  & $\log(A)$  & $\sigma$  & N & $\log(A)$  & $\sigma$   &[X/H] & $\sigma$  & N& $\log(A)$ &$\sigma$& [X/H]   & $\sigma$& [A-B] & $\sigma$ &  N & $\log(A)$& $\sigma$&[X/H]&  $\sigma$& N&$\log(A)$ & $\sigma$& [X/H]    & $\sigma$&   [A-B]        & $\sigma$&   \\
			\hline
			Li\ione & 1  &-10.93~ & 0.10~~& 1 & -10.58~ & 0.06~   & 0.35~ & 0.10~ & 1&  -11.29~ & 0.20~   &-0.36~&  0.10~&~0.71~ & 0.10~&   1&  -10.64~&  0.07~&    0.29~& 0.07~  &  1&  -11.41~& 0.20~ &   -0.48~&  0.20~&   0.77~&  0.10~&     \\
			C(mol) $^1$ &    & -3.622 & 0.049 &   &  -3.570 & 0.029   & 0.052 & 0.029 &  &   -3.554 & 0.032   &~0.054&  0.032&-0.016 & 0.030&    &   -3.522&  0.031&    0.100&  0.030 &   &   -3.544&  0.028&   ~0.078&  0.030&   0.022&  0.030& ~~40\\
			C\ione  & 7  & -3.601 & 0.027 & 7 &  -3.560 & 0.037   & 0.041 & 0.025 & 7&   -3.564 & 0.037   &~0.037&  0.017&~0.004 & 0.015&   7&   -3.598&  0.048&    0.003&  0.040 &  7&   -3.606&  0.046&   -0.005&  0.038&  ~0.009&  0.010& ~~40\\
			%N\ione  & 2  & -4.030 & 0.024 & 2 &  -4.028 & 0.009   & 0.002 & 0.033 & 2&   -4.039 & 0.002   &-0.009&  0.022&~0.011 & 0.011&   2&   -4.028&  0.010&    0.002&  0.014 &  2&   -4.032&  0.011&   -0.002&  0.013&  ~0.004&  0.001& ~123\\
			N\ione  & 2  & -4.062 & 0.033 & 2 &  -4.063 & 0.009   &-0.002 & 0.024 & 2&   -4.076 & 0.017   &-0.015&  0.016&~0.013 & 0.008&   2&   -4.066&  0.022&   -0.005&  0.054 &  2&   -4.069&  0.029&   -0.007&  0.061&  ~0.002&  0.007& ~123\\
			O\ione  & 5  & -3.332 & 0.086 & 5 &  -3.300 & 0.060   & 0.032 & 0.030 & 4&   -3.289 & 0.063   &~0.043&  0.025&-0.011 & 0.007&   5&   -3.337&  0.060&   -0.006&  0.028 &  5&   -3.343&  0.073&   -0.011&  0.028&  ~0.005&  0.020& ~180\\
			Na\ione & 6  & -5.888 & 0.060 & 6 &  -5.780 & 0.052   & 0.108 & 0.014 & 6&   -5.799 & 0.053   &~0.089&  0.012&~0.019 & 0.011&   6&   -5.769&  0.056&    0.119&  0.013 &  6&   -5.779&  0.058&   ~0.109&  0.014&  ~0.011&  0.010& ~958\\
			Mg\ione & 4  & -4.442 & 0.099 & 4 &  -4.337 & 0.103   & 0.106 & 0.023 & 5&   -4.364 & 0.106   &~0.079&  0.018&~0.027 & 0.007&   4&   -4.295&  0.101&    0.147&  0.019 &  4&   -4.306&  0.107&   ~0.137&  0.027&  ~0.010&  0.009& 1336\\
			Al\ione & 6  & -5.546 & 0.028 & 6 &  -5.404 & 0.037   & 0.142 & 0.015 & 6&   -5.418 & 0.036   &~0.128&  0.020&~0.014 & 0.010&   6&   -5.367&  0.040&    0.179&  0.020 &  6&   -5.380&  0.035&   ~0.166&  0.014&  ~0.013&  0.008& 1653\\
			Si\ione &18  & -4.538 & 0.086 &18 &  -4.440 & 0.085   & 0.098 & 0.016 &18&   -4.467 & 0.090   &~0.071&  0.024&~0.027 & 0.015&  18&   -4.437&  0.084&    0.102&  0.012 & 18&   -4.459&  0.086&   ~0.079&  0.018&  ~0.023&  0.012& 1310\\
			S\ione  & 7  & -4.803 & 0.025 & 7 &  -4.747 & 0.036   & 0.056 & 0.034 & 7&   -4.765 & 0.035   &~0.038&  0.031&~0.017 & 0.006&   7&   -4.762&  0.028&    0.041&  0.024 &  7&   -4.780&  0.030&   ~0.023&  0.031&  ~0.018&  0.018& ~664\\
			%S\ione  & 6  & -4.800 & 0.026 & 6 &  -4.751 & 0.037   & 0.049 & 0.033 & 6&   -4.768 & 0.037   &~0.033&  0.031&~0.016 & 0.006&   6&   -4.764&  0.031&    0.037&  0.023 &  6&   -4.781&  0.032&   ~0.020&  0.033&  ~0.017&  0.019& ~664\\
			K\ione  & 3  & -6.976 & 0.102 & 3 &  -6.873 & 0.044   & 0.103 & 0.060 & 3&   -6.883 & 0.050   &~0.094&  0.056&~0.010 & 0.007&   3&   -6.842&  0.083&    0.134&  0.035 &  3&   -6.849&  0.072&   ~0.128&  0.050&  ~0.007&  0.016& 1006\\
			Ca\ione &13  & -5.717 & 0.050 &13 &  -5.619 & 0.053   & 0.097 & 0.013 &13&   -5.646 & 0.052   &~0.070&  0.011&~0.027 & 0.008&  13&   -5.586&  0.055&    0.131&  0.010 & 13&   -5.595&  0.054&   ~0.122&  0.014&  ~0.009&  0.009& 1517\\
			Sc\ii $^1$  & 6  & -8.873 & 0.048 & 6 &  -8.732 & 0.059   & 0.141 & 0.015 & 7&   -8.760 & 0.059   &~0.113&  0.013&~0.028 & 0.005&   6&   -8.710&  0.068&    0.163&  0.028 &  6&   -8.741&  0.066&   ~0.132&  0.023&  ~0.031&  0.007& 1659\\
			Ti\ione &35  & -7.101 & 0.054 &35 &  -6.993 & 0.057   & 0.108 & 0.025 &35&   -7.023 & 0.061   &~0.078&  0.027&~0.029 & 0.017&  35&   -6.957&  0.064&    0.144&  0.029 & 35&   -6.956&  0.067&   ~0.145&  0.030&  -0.001&  0.018& 1582\\
			Ti\ii   & 9  & -7.072 & 0.047 & 9 &  -6.937 & 0.053   & 0.135 & 0.020 & 9&   -6.962 & 0.058   &~0.110&  0.025&~0.025 & 0.012&   9&   -6.935&  0.049&    0.137&  0.014 &  9&   -6.956&  0.061&   ~0.116&  0.024&  ~0.021&  0.018& 1582\\
			V\ione $^1$ &19  & -8.115 & 0.052 &19 &  -8.009 & 0.030   & 0.107 & 0.028 &19&   -8.040 & 0.031   &~0.076&  0.030&~0.031 & 0.016&  19&   -7.974&  0.037&    0.142&  0.035 & 19&   -7.990&  0.032&   ~0.125&  0.039&  ~0.017&  0.013& 1429\\
			%LTE Cr\ione &19  & -6.387 & 0.076 &19 &  -6.309 & 0.079   & 0.078 & 0.022 &20&   -6.333 & 0.073   &~0.055&  0.022&~0.023 & 0.012&  19&   -6.271&  0.082&    0.116&  0.029 & 19&   -6.282&  0.077&   ~0.106&  0.029&  ~0.011&  0.012& 1296\\
			Cr\ione &19  & -6.321 & 0.078 &19 &  -6.242 & 0.081   & 0.079 & 0.022 &20&   -6.265 & 0.075   &~0.055&  0.022&~0.024 & 0.012&  19&   -6.209&  0.080&    0.111&  0.029 & 19&   -6.221&  0.076&   ~0.099&  0.030&  ~0.012&  0.012& 1296\\
			Cr\ii   & 8  & -6.349 & 0.059 & 8 &  -6.256 & 0.071   & 0.093 & 0.018 & 8&   -6.281 & 0.067   &~0.068&  0.016&~0.025 & 0.013&   8&   -6.271&  0.067&    0.078&  0.017 &  8&   -6.295&  0.068&   ~0.054&  0.015&  ~0.024&  0.013& 1296\\
			%LTE Mn\ione &12  & -6.601 & 0.073 &12 &  -6.513 & 0.071   & 0.089 & 0.009 &14&   -6.539 & 0.070   &~0.062&  0.010&~0.026 & 0.009&  12&   -6.485&  0.062&    0.116&  0.016 & 12&   -6.496&  0.062&   ~0.105&  0.015&  ~0.011&  0.008& 1158\\
			Mn\ione &12  & -6.570 & 0.070 &12 &  -6.478 & 0.068   & 0.092 & 0.009 &14&   -6.508 & 0.067   &~0.062&  0.010&~0.030 & 0.009&  12&   -6.450&  0.059&    0.120&  0.017 & 12&   -6.464&  0.059&   ~0.107&  0.015&  ~0.014&  0.009& 1158\\
			Fe\ione &73  & -4.530 & 0.067 &73 &  -4.434 & 0.070   & 0.096 & 0.017 &73&   -4.462 & 0.073   &~0.068&  0.016&~0.028 & 0.012&  73&   -4.392&  0.069&    0.138&  0.022 & 73&   -4.403&  0.073&   ~0.127&  0.023&  ~0.011&  0.012& 1334\\
			Fe\ii   &23  & -4.538 & 0.049 &23 &  -4.441 & 0.052   & 0.097 & 0.023 &23&   -4.461 & 0.047   &~0.077&  0.023&~0.020 & 0.012&  23&   -4.444&  0.053&    0.094&  0.022 & 23&   -4.461&  0.050&   ~0.077&  0.021&  ~0.018&  0.012& 1334\\
			%LTE Co\ione &12  & -7.114 & 0.070 &12 &  -7.027 & 0.073   & 0.088 & 0.022 &12&   -7.053 & 0.064   &~0.061&  0.022&~0.027 & 0.015&  12&   -7.001&  0.075&    0.113&  0.026 & 12&   -7.012&  0.069&   ~0.103&  0.027&  ~0.011&  0.016& 1352\\
			Co\ione &12  & -7.034 & 0.076 &12 &  -6.947 & 0.079   & 0.087 & 0.022 &12&   -6.978 & 0.070   &~0.055&  0.022&~0.031 & 0.015&  12&   -6.922&  0.082&    0.111&  0.026 & 12&   -6.936&  0.076&   ~0.097&  0.026&  ~0.014&  0.016& 1352\\
			Ni\ione $^1$ &18  & -5.813 & 0.071 &18 &  -5.715 & 0.071   & 0.099 & 0.021 &18&   -5.744 & 0.072   &~0.070&  0.017&~0.029 & 0.009&  18&   -5.688&  0.076&    0.126&  0.031 & 18&   -5.706&  0.077&   ~0.108&  0.028&  ~0.018&  0.009& 1353\\
			Cu\ione $^1$ & 6  & -7.865 & 0.047 & 6 &  -7.749 & 0.053   & 0.116 & 0.021 & 6&   -7.771 & 0.046   &~0.094&  0.011&~0.022 & 0.014&   6&   -7.723&  0.063&    0.142&  0.030 &  6&   -7.730&  0.060&   ~0.135&  0.025&  ~0.007&  0.016& 1037\\
			Zn\ione & 3  & -7.505 & 0.006 & 3 &  -7.382 & 0.006   & 0.123 & 0.008 & 3&   -7.393 & 0.013   &~0.112&  0.007&~0.011 & 0.015&   3&   -7.352&  0.028&    0.153&  0.022 &  3&   -7.358&  0.031&   ~0.147&  0.024&  ~0.005&  0.010& ~726\\
			%Rb\ione & 2  & -9.498 & 0.027 & 2 &  -9.364 & 0.032   & 0.134 & 0.004 & 1&   -9.393 & 0.018   &~0.106&  0.010&~0.028 & 0.014&   2&   -9.340&  0.029&    0.158&  0.001 &  2&   -9.355&  0.016&   ~0.143&  0.011&  ~0.015&  0.012& ~800\\
			Rb\ione & 2  & -9.618 & 0.027 & 2 &  -9.494 & 0.032   & 0.124 & 0.004 & 1&   -9.523 & 0.018   &~0.096&  0.010&~0.028 & 0.014&   2&   -9.470&  0.029&    0.148&  0.001 &  2&   -9.485&  0.016&   ~0.133&  0.011&  ~0.015&  0.012& ~800\\
			Y\ii $^1$   & 7  & -9.870 & 0.050 & 7 &  -9.819 & 0.051   & 0.051 & 0.015 & 7&   -9.837 & 0.059   &~0.033&  0.013&~0.017 & 0.013&   7&   -9.822&  0.061&    0.048&  0.045 &  7&   -9.831&  0.072&   ~0.039&  0.051&  ~0.010&  0.015& 1659\\
			Zr\ii   & 3  & -9.368 & 0.032 & 3 &  -9.315 & 0.039   & 0.053 & 0.023 & 3&   -9.322 & 0.050   &~0.046&  0.023&~0.007 & 0.015&   3&   -9.294&  0.037&    0.074&  0.020 &  3&   -9.295&  0.054&   ~0.073&  0.028&  ~0.001&  0.018& 1741\\
			Ba\ii   & 4  & -9.835 & 0.038 & 4 &  -9.764 & 0.036   & 0.071 & 0.012 & 4&   -9.797 & 0.043   &~0.038&  0.006&~0.032 & 0.016&   4&   -9.751&  0.034&    0.084&  0.013 &  4&   -9.776&  0.037&   ~0.060&  0.007&  ~0.024&  0.015& 1455\\
			La\ii $^1$  & 2  &-10.947 & 0.018 & 2 & -10.898 & 0.016   & 0.049 & 0.002 & 2&  -10.915 & 0.010   &~0.032&  0.007&~0.018 & 0.005&   2&  -10.889&  0.024&    0.057&  0.007 &  2&  -10.913&  0.030&   ~0.033&  0.013&  ~0.024&  0.006& 1578\\
			Ce\ii $^1$  & 3  &-10.463 & 0.013 & 3 & -10.375 & 0.022   & 0.088 & 0.029 & 3&  -10.396 & 0.024   &~0.067&  0.035&~0.021 & 0.016&   3&  -10.378&  0.050&    0.085&  0.060 &  3&  -10.384&  0.033&   ~0.079&  0.045&  ~0.006&  0.021& 1478\\
			Nd\ii $^1$  & 6  &-10.643 & 0.040 & 6 & -10.578 & 0.034   & 0.065 & 0.042 & 6&  -10.598 & 0.032   &~0.045&  0.037&~0.020 & 0.010&   6&  -10.574&  0.041&    0.069&  0.054 &  6&  -10.577&  0.033&   ~0.066&  0.038&  ~0.003&  0.019& 1602\\
			Sm\ii $^1$  & 5  &-11.101 & 0.103 & 5 & -11.069 & 0.053   & 0.032 & 0.086 & 5&  -11.072 & 0.067   &~0.029&  0.094&~0.003 & 0.017&   5&  -11.057&  0.085&    0.044&  0.065 &  5&  -11.048&  0.088&   ~0.053&  0.075&  -0.009&  0.019& 1590\\
			\hline                                                                                                         
			\multicolumn{29}{l}{{\it Note.} $^1$ LTE abundances.} \\ 
		\end{tabular}                                                                      
	\end{small}
	%NLTE correction -0.13~dex is applied to all Rb mean abundances according to \citet{2020AstL...46..541K}.
%\end{table}
\end{sidewaystable*}

%\end{landscape}

\begin{table*}
\caption{Average NLTE abundances of HD~219542~A and HD~219542~B.} 
\label{HD219542-abun}
%	\begin{small}
	\tabcolsep=2.2pt
	\begin{tabular}{l rrc @{\hspace{10pt}} cc @{\hspace{20pt}} rrc @{\hspace{10pt}} cc  @{\hspace{20pt}} cc}
		\hline
		& \multicolumn{5}{c}{HD~219542~A}& \multicolumn{5}{c}{HD~219542~B}& & \\
		Elem    & N  & $\log(A)$  & $\sigma$  &[X/H] & $\sigma$  & N & $\log(A)$  & $\sigma$   &[X/H] & $\sigma$  & [A-B] & $\sigma$ \\
		\hline
		Li\ione & 1&   -9.84~~&0.05~ &    1.09~~& 0.11~~&  1&  -10.86~& 0.10~~&   0.07~~&  0.14~~& ~1.02~ & 0.11~\\
		C(mol) $^1$ &  &   -3.54~~&0.05~~&    0.07~~& 0.07~~&   &  -3.53~~& 0.04~~&   0.08~~& 0.06~~& -0.01~~& 0.06~~\\
		C\ione  & 5&   -3.500&  0.021&    0.098&  0.031 &  5&   -3.508&  0.028&    0.090&  0.047&  ~0.007&  0.019\\
		O\ione  & 2&   -3.228&  0.026&    0.015&  0.043 &  2&   -3.220&  0.018&    0.022&  0.051&  -0.007&  0.008\\  
		Na\ione & 4&   -5.748&  0.042&    0.109&  0.012 &  4&   -5.729&  0.051&    0.128&  0.003&  -0.020&  0.013\\
		Mg\ione & 3&   -4.366&  0.098&    0.076&  0.017 &  3&   -4.342&  0.086&    0.100&  0.031&  -0.024&  0.020\\
		Al\ione & 2&   -5.473&  0.002&    0.083&  0.014 &  2&   -5.457&  0.002&    0.100&  0.015&  -0.016&  0.001\\
		Si\ione &17&   -4.418&  0.090&    0.124&  0.025 & 17&   -4.407&  0.090&    0.135&  0.022&  -0.011&  0.011\\
		S\ione  & 2&   -4.689&  0.005&    0.114&  0.032 &  4&   -4.697&  0.009&    0.107&  0.036&  ~0.007&  0.004\\  
		Ca\ione &13&   -5.594&  0.059&    0.123&  0.039 & 13&   -5.570&  0.048&    0.147&  0.027&  -0.024&  0.025\\
		Sc\ii  $^1$  & 6&   -8.731&  0.072&    0.142&  0.058 &  6&   -8.736&  0.081&    0.137&  0.072&  ~0.005&  0.018\\  
		Ti\ione &33&   -7.025&  0.052&    0.072&  0.034 & 33&   -6.981&  0.057&    0.116&  0.034&  -0.044&  0.014\\
		Ti\ii   & 9&   -6.962&  0.056&    0.110&  0.044 &  9&   -6.965&  0.058&    0.108&  0.045&  ~0.003&  0.011\\
		V\ione $^1$ &16&   -8.031&  0.034&    0.081&  0.040 & 16&   -7.975&  0.032&    0.137&  0.045&  -0.056&  0.023\\
		%LTE			Cr\ione &17&   -6.275&  0.056&    0.116&  0.039 & 17&   -6.248&  0.065&    0.143&  0.025&  -0.027&  0.030\\ 
		Cr\ione &17&   -6.208&  0.055&    0.111&  0.039 & 17&   -6.178&  0.069&    0.141&  0.026&  -0.030&  0.031\\ 
		Cr\ii   & 8&   -6.221&  0.052&    0.127&  0.020 &  8&   -6.242&  0.073&    0.107&  0.040&  ~0.021&  0.033\\  
		%LTE			Mn\ione &12&   -6.510&  0.079&    0.091&  0.036 & 12&   -6.470&  0.073&    0.131&  0.028&  -0.040&  0.018\\  
		Mn\ione &12&   -6.475&  0.077&    0.095&  0.037 & 12&   -6.442&  0.070&    0.128&  0.028&  -0.033&  0.019\\  
		Fe\ione &69&   -4.441&  0.071&    0.085&  0.053 & 69&   -4.416&  0.078&    0.111&  0.060&  -0.026&  0.025\\  
		Fe\ii   &23&   -4.416&  0.063&    0.127&  0.048 & 23&   -4.431&  0.072&    0.113&  0.045&  ~0.015&  0.033\\  
		%LTE			Co\ione &12&   -7.000&  0.081&    0.114&  0.046 & 12&   -6.995&  0.075&    0.119&  0.028&  -0.004&  0.036\\
		Co\ione &12&   -9.918&  0.089&    0.115&  0.046 & 12&   -6.930&  0.080&    0.104&  0.027&  ~0.011&  0.036\\
		Ni\ione $^1$ &15&   -5.701&  0.059&    0.117&  0.034 & 15&   -5.679&  0.068&    0.140&  0.032&  -0.023&  0.016\\
		Cu\ione $^1$ & 3&   -7.831&  0.018&    0.078&  0.035 &  3&   -7.808&  0.005&    0.101&  0.022&  -0.023&  0.013\\
		Zn\ione & 2&   -7.416&  0.006&    0.084&  0.005 &  2&   -7.396&  0.007&    0.104&  0.006&  -0.020&  0.002\\
		Y\ii $^1$   & 7&   -9.775&  0.065&    0.095&  0.027 &  7&   -9.782&  0.070&    0.088&  0.031&  ~0.007&  0.026\\
		Zr\ii   & 3&   -9.223&  0.038&    0.145&  0.021 &  3&   -9.206&  0.042&    0.162&  0.010&  -0.017&  0.021\\
		Ba\ii   & 4&   -9.727&  0.022&    0.108&  0.042 &  4&   -9.691&  0.009&    0.144&  0.039&  -0.036&  0.014\\
		Ce\ii $^1$  & 2&  -10.313&  0.008&    0.145&  0.005 &  2&  -10.285&  0.022&    0.173&  0.009&  -0.028&  0.014\\
		\hline                                                                                                                    
		\multicolumn{13}{l}{{\it Note.} $^1$ LTE abundances.} \\ 
	\end{tabular}                                                                      
	%	\end{small}
\end{table*}

\begin{figure*}           
	\includegraphics[width=0.95\textwidth,clip]{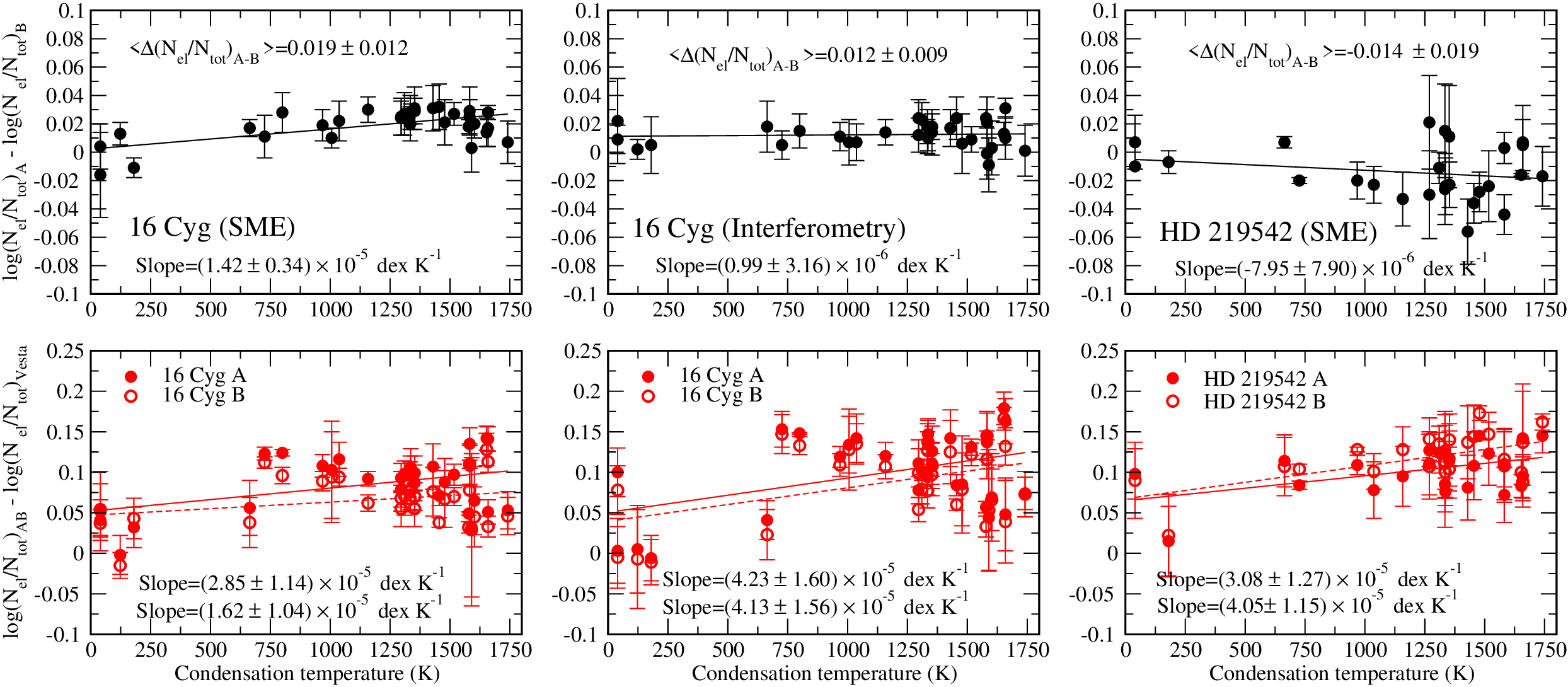}
	\caption{Abundance differences between the A and B components of 16~Cyg and HD~219542 (top panels, black circles) and relative-to-solar abundances of the stars (bottom panels) versus the dust condensation temperatures. The results based on the \SME\ atmospheric parameters for 16~Cyg and HD~219542 are shown in the left and right panels. The central panels display the results based on the interferometric atmospheric parameters for 16~Cyg. 
Linear regressions are shown by solid black lines in the top panels and by solid (primaries) and dashed (secondaries) red lines in the bottom panels.}
	\label{abund-CT}
\end{figure*}

% It gives a credit to the results of our line-by-line differential abundance analysis. 
Figure~\ref{abund-CT} displays the abundance differences between the A and B components of 16~Cyg and HD~219542 and the relative-to-solar abundances of the stars as a function of the dust condensation temperature. Our findings can be formulated as follows.
\begin{itemize}
	\item In both binary systems, independent of the presence or the absence of the planet, the A and B components do not reveal significant abundance discrepancies, with A -- B = 0.019$\pm$0.012~dex and $-0.014\pm$0.019~dex for 16~Cyg and HD~219542, respectively.
	 This provides a further evidence for a negligible effect of the giant planet formation on atmospheric abundances of the host star \citep[see, for instance,][]{2014MNRAS.442L..51L, 2018A&A...614A.138L}. Although a slow-growing trend with \CT\ and a hint of decreasing (A -- B) for \CT\ $>$ 1400~K are seen for 16~Cyg in the case of using the \SME\ atmospheric parameters. A fairly flat distribution of (A -- B) with \CT\ is obtained for the interferometric atmospheric parameters. In Sect.~\ref{sect:Discussion}, we discuss an influence of atmospheric parameters on the abundance trends.
\item 16~Cyg and HD~219542 reveal a different behaviour of [X/H] with \CT. For both components of HD~219542, the observed trends are quasi-linear and similar to the trend of [X/H] with \CT\ for the solar analogues without detected planets \citep[see Fig.~5 in][]{2009ApJ...704L..66M}.
Both components of 16~Cyg reveal a meaningful step-like change in [X/H] at \CT\ $\approx$ 750~K, then a decline with increasing \CT\ up to $\approx$ 1300~K, and splitting into two trends for the higher \CT. A complex behaviour of [X/H] with \CT\ is even more evident for the interferometric atmospheric parameters.
%The [X/H] trends for 16~Cyg reveal a complex structure, 
See Section~\ref{sect:Discussion} for a discussion.
%meaningful step at \CT$\approx$750~K that may indicate a specific chemical composition of the original cloud.
\end{itemize}

%	Our analysis did not reveal significant abundance difference between the components of both binary systems, although, 
%	a small trend with \CT\ is seen in differential abundances (A-B) for 16~Cyg, but it differs from the famous trend 'Sun-solar twins' discovered in \citet{2009ApJ...704L..66M} and interpreted as possible effect of solar system planet formation. 
%	However, a certain difference between two systems exists when we compare the relative-to-solar abundances of the components as a function of the dust condensation temperature. In HD~219542 the observed trends are quasi-linear and similar to the trends in the differences between the Sun and the mean values of the solar analogs without detected planets as a function of \CT\ \citep[Fig.5 in][]{2009ApJ...704L..66M}, while in 16~Cyg the trend has a complex structure (more details will be given in Section~\ref{sect:Discussion}). 
	%or significant trends with the condensation temperature.
%	Two wide pairs studied here differ by the presence of a detected close-in giant planet in one of them, 16 Cyg. Identical chemistry of the components in both pairs allows us to provide a further support to the conclusion that the giant planet formation has a negligible effect on the atmosphere of the host star \citep[see, for instance,][]{2014MNRAS.442L..51L, 2018A&A...614A.138L}. }        

\section{Discussion}\label{sect:Discussion}

\subsection{Li abundances and age estimates}

Lithium is of special interest because its deficiency was found in the planet-host atmospheres \citep{1997AJ....113.1871K, 2000AJ....119..390G}. The Li deficiency in 16~Cyg~B was confirmed by a number of studies (see references in Table~\ref{param_published}), while the atmosphere of 16~Cyg~A is more Li rich compared with the solar one. The Li abundance difference between the A and B components in 16~Cyg reaches 0.7 -- 0.8~dex \citep[see Table~7 in ][]{2019AA...628A.126M}. Even larger difference in Li abundance between the A and B components, of $>$1~dex, was derived for HD~219542 \citep{2001AA...377..123G, 2003PASJ...55.1005S}. 

%NLTE Li abundances were derived based on line profile fitting of Li\ione\, $\lambda$6707.8~\AA\, feature in \SME taking into account the isotopic and hyperfine splittings as well as grids of departure coefficients calculated according \citet{2009A&A...503..541L}. 
%The final Li abundances are given in Table~\ref{Vesta-16Cyg} (Vesta, 16~CygAB) and in Table~\ref{HD219542-abun} (HD~219542AB).
 
For 16~Cyg, we obtained the Li abundance difference [A-B] = 0.71$\pm$0.10 and 0.77$\pm$0.10 for two sets of atmospheric parameters. [A-B] = 1.02~dex for the HD~219542 system. Our results agree with the literature data in the following aspects: (i) the hotter components of each system have supersolar Li abundance; (ii) the planet-host star 16~Cyg~B is deficient in Li; (iii) HD~219542~B without detected planet has close-to-solar Li abundance.  

\begin{figure*}           
	\includegraphics[width=0.85\textwidth,clip]{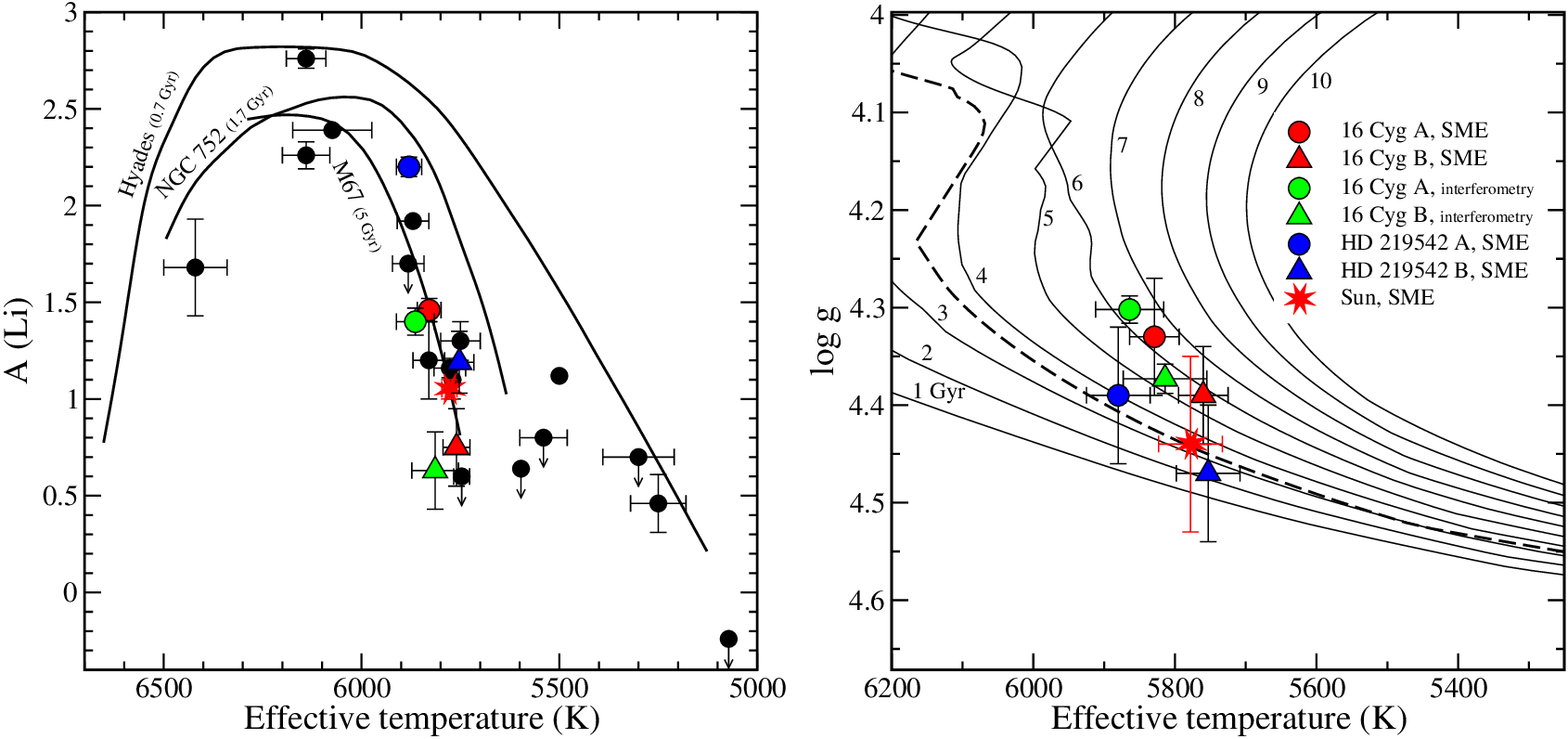}
	\caption{Left panel: Li abundances of 16~Cyg (red and green colour symbols correspond to the \SME\ and interferometric atmospheric parameters, respectively) and HD~219542 (blue colour symbols) components compared with the literature data for the planet-host stars (black circles) and the Li abundance versus \te\ trends for open clusters of different ages (solid curves). See text for references. The primaries are shown by circles, the secondaries are shown by triangles. The solar Li abundance is shown by red asterisk. 
%		displayed on Fig.2 of \citet{2000MNRAS.316L..35R} where black filled circles indicate planet-host stars. Red, green, and blue circles show 16~Cyg (\SME\ parameters), 16~Cyg (interferometric parameters), and HD~219542 (\SME\ parameters). The solar Li abundance is shown by red asteriscs. Trends in Li abundance versus \te\ for open clusters of different ages (solid lines) are taken from \citet{1988ApJ...334..734H} (NGC~752, $t$=1.7~Gyr), and from \citet{1995ApJ...453..819R} (Hyades, $t$=0.7~Gyr; M67, $t$=5~Gyr). \newline
	Right panel: Position of the Sun, 16~Cyg~AB, and HD~219542~AB on the MESA isochrones calculated for the solar abundances (age = 4.4~Gyr, dashed curve) and [M/H] = +0.1~dex (solid curves).} \label{Li}
\end{figure*}

The Li abundance difference as well as the overall metallicity difference between the A and B components of 16~Cyg were attributed to an accretion of planetary material onto 16~Cyg~A or the planet engulfment event \citep{2001ApJ...553..405L, 2019AA...628A.126M}. 
However, there may be a different and natural explanation of the difference in Li abundance between the components. First, analysis of stars in the open clusters shows a rapid decrease in the Li abundance with decreasing effective temperature \citep[see Li abundance trends in Fig.~2 of][]{2000MNRAS.316L..35R}. For open clusters of $t$ = 1.7 -- 5~Gyr, the Li abundance drops from $\eps{Li}$ = 2.5 to 0.6 within a narrow temperature region of $\sim$200-300~K. Here, the abundance scale is used where $\eps{H}$ = 12. Therefore, for both 16~Cyg and HD~219542, the Li abundance difference between the components can be explained by their temperature difference. We produced the plot similar to Fig.~2 in \citet{2000MNRAS.316L..35R}. In the left panel of Fig.~\ref{Li}, the Li abundance versus \te\ trends for open clusters of different ages were taken from \citet[][NGC~752, $t$ = 1.7~Gyr]{1988ApJ...334..734H} and \citet[][Hyades, $t$ = 0.7~Gyr; M67, $t$ = 5~Gyr]{1995ApJ...453..819R}. The Li abundances of the planet-host stars were taken from Table~1 of \citet{2000MNRAS.316L..35R}.
   
Second, for the solar twin stars (\te\ = 5700-5860~K), the Li abundance drops with increasing stellar age \citep{2016A&A...587A.100C}, namely, from $\eps{Li}$ = 2.2 to 0.6 over an age interval of 1$\le{t}\le$8~Gyr. We estimated ages of both systems using their \logg\ and \te\ and the isochrones from the single-star MESA (Modules for Experiments in Stellar Astrophysics) theoretical stellar evolution model grid\footnote{\url{http://waps.cfa.harvard.edu/MIST/}} \citep{2016ApJ...823..102C}. We checked the isochrone of the solar chemical composition and age of 4.4~Gyr and the isochrones of [M/H] = +0.1~dex (right panel of Fig.~\ref{Li}). We estimated $t$ = 6$\pm$1~Gyr for the 16~Cyg system and $t$ = 3$\pm$1~Gyr for HD~219542. Our estimate for 16~Cyg agrees within the error bars with the ages, $t$(A) = 6.80$\pm$0.4 and $t$(B) = 6.90$\pm$0.28~Gyr, derived from the astroseismic modelling of \citet{2017ApJ...851...80B}. 

We conclude that the effective temperature difference may explain the Li abundance difference between the components, without 
involvement a hypothesis of planetary material accretion onto one of the components of binary system. The age difference between the two systems may explain the higher Li abundance of the HD~219542 components compared to those of 16~Cyg. 

%the  Figure~\ref{Li} shows that 

\subsection{Influence of uncertainties in atmospheric parameters on differential abundances}

\begin{figure*}           
	\includegraphics[width=0.9\textwidth,clip]{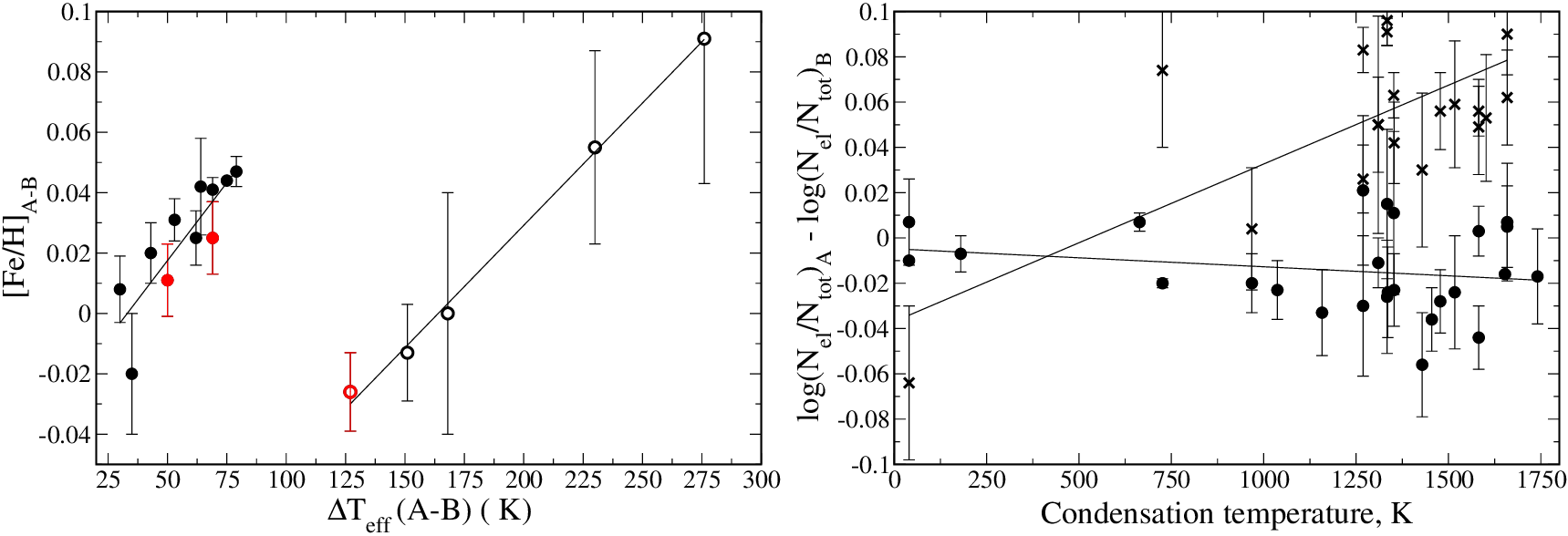}
	\caption{Left panel: Metallicity difference between the components of 16~Cyg (filled circles) and HD~219542 (open circles) depending on the effective temperature difference. The literature data from Table~\ref{param_published} are shown by black symbols and our results by red symbols.
% \newline
		Right panel: Abundance differences between the components of HD~219522 from this study (filled circles, $\Delta$\te (A-B) = 127~K) and \citet{2001AA...377..123G} (crosses, $\Delta$\te (A-B) = 276~K).}\label{DeltaA-DeltaT}
\end{figure*}

\citet{2015ApJ...801L..10T} investigated how the differential abundances in the planet-host binary system XO-2 change with the changes in atmospheric parameters of the components, choice of spectral lines, and methods of spectrum analysis.
It was shown that the \te\ difference plays the major role in shaping the relative abundance versus \CT\ trend that is usually considered as an indicator of the planet formation effects. The slope of the trend may change from positive to negative depending on the size of the effective temperature difference between the components \citep[see Fig.~1 in][]{2015ApJ...801L..10T}. Their conclusion is supported by our results. Figure~\ref{DeltaA-DeltaT} (left panel) displays the Fe abundance difference as a function of the \te\ difference between the components of 16~Cyg and HD~219542. It is evident that these two parameters reveal a linear correlation.
% both systems on  and see practically between. 
The right panel of Fig.~\ref{DeltaA-DeltaT} shows how a slope of the [Fe/H]$_{\rm A-B}$ = [Fe/H]$_{\rm A}$ -- [Fe/H]$_{\rm B}$ versus \CT\ trend for HD~219542 changes when moving to the largest temperature difference (276~K) between the components derived by \citet{2001AA...377..123G}. We conclude from this that the abundance differences derived in some previous studies for the HD~219542~A and B can be removed by revising their atmospheric parameters.
% The results of the comparison mean that at least in binary system HD~219542 it's possible to find a reasonable combination of the atmospheric parameters of both components which cancel any abundance difference derived in some previous analyses. 
 
 The situation is slightly different for the 16~Cyg system. Except one case in Table~\ref{param_published} with a negative [Fe/H] difference, all the other analyses obtain that 16~Cyg~A is marginally more metal-rich than the planet-host star 16~Cyg~B, by 0.02 to 0.047~dex. This study obtained the average abundance differences [A-B] = 0.019~dex and 0.012~dex, when using the \SME\ and interferometric atmospheric parameters, respectively. 
 
Using the code\footnote{\url{https://github.com/ramstojh/terra}} by \citet{2016A&A...589A..65G}, we estimated the mass of rocky material needed to explain the abundance difference between the A and B components of 16~Cyg. First, the code determines the convective mass of the component A for a given atmospheric parameters. Then, for each chemical element, its mass fraction is calculated by adding 
%the amount of 
a mixture from the meteoritic \citep[chondrites,][]{1988RSPTA.325..535W} and terrestrial \citep{2001E&PSL.185...49A} abundance patterns to that for the B component. The latter is approximated by the solar chemical composition scaled with the overall metallicity of the B component. For more details, see the Appendix in \citet{2016A&A...589A..65G}. We obtained that the slightly higher metallicity of 16~Cyg~A compared with 16~Cyg~B could be caused by an engulfment of the $\sim$2~Earth mass planet composed of the meteoritic (chondrites) like material.
 
% Terra uses chondritic material from \citet{{1988RSPTA.325..535W} and terrestrial one from \citet{2001E&PSL.185...49A}    

\subsection{Trend of abundance differences with the condensation temperature}
		
\begin{figure}           
	\includegraphics[width=0.45\textwidth,clip]{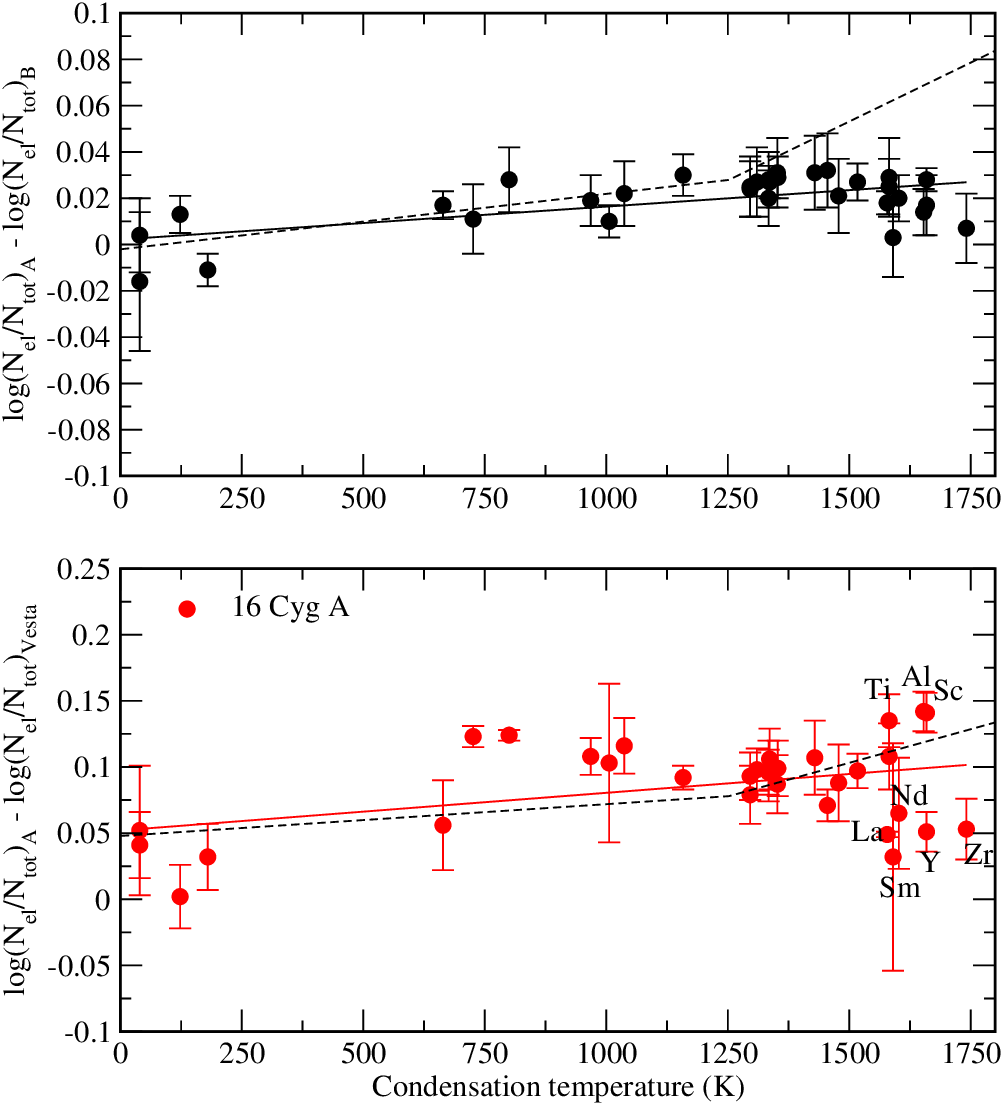}
	\caption{Abundance differences between the components of 16~Cyg (top panel) and between 16~Cyg~A and the Sun (bottom panel) as function of \CT. Linear regressions are shown by the solid lines. The 'solar twins -- Sun' abundance trend of \citet{2009ApJ...704L..66M} is displayed by the dashed curve. It is shifted in Y-axe by 0.05~dex in the top panel, for a clearer comparison with our data, and by 0.10~dex in the bottom panel, in line with the metallicity of 16~Cyg~A.}\label{Gust}
\end{figure}

\citet{2009ApJ...704L..66M} have discovered the 'Sun -- solar twins' trend, which is interpreted as a possible effect of the solar system planet formation. Their trend transformed to the 'solar twins -- Sun' one is reproduced in Fig.~\ref{Gust} together with our abundance trends for 16~Cyg. The [A-B] abundance differences for 16~Cyg are expected to follow the 'solar twins -- Sun' trend because 16~Cyg~B has a planet like the Sun and 16~Cyg~A is a solar twin. Both trends in the top panel of Fig.~\ref{Gust} reveal a common behaviour for \CT\ $<$ 1300~K, however, there is a hint of decreasing [A-B] for \CT\ $>$ 1400~K, while, in the same \CT\ range, the solar twins -- Sun differences grow with a bigger slope compared with that for the lower temperatures.
% slow-growing trend with \CT, are different 

%A behavior of the [X/H] abundances is complex for both components of 16~Cyg (Fig.~\ref{abund-CT}, left and middle bottom panels). In 
Figure~\ref{Gust} (bottom panel) shows that the [X/H] abundances of 16~Cyg~A reveal a different behaviour compared with the 'solar twins -- Sun' trend. Note that both components of 16~Cyg independent of using either the \SME\ or the interferometric atmospheric parameters reveal very similar trends of [X/H] with \CT. Neither a step-like change in [X/H] at \CT\ $\approx$ 750~K, nor splitting into two trends for the highest \CT\ were detected so far in any of the solar twins.  

The 16~Cyg binary system is older than the Sun, and one could expect that it has the lower abundances compared with the solar ones, for example, by 0.07~dex for C and O, by 0.10~dex for S, and by 0.12~dex for Fe, according to the Galactic chemical evolution model of \citet{2020ApJ...900..179K}. However, 16~Cyg is enhanced in all the elements relative to the Sun. Therefore, the observed features in the [X/H] abundances cannot be explained by the Galactic chemical evolution effects. 
%It differs from the 'Sun -- solar twins' trends discovered by \citet{2009ApJ...704L..66M} and interpreted as possible effect of solar system planet formation
%We clearly see a meaningful step at \CT$\approx$750~K for both components, then a decline up to \CT $\approx$1300~K following by the splitting of the trend into two parts (bottom panel in Fig.~\ref{Gust}). 
The similarity of the abundance trends for both components indicates a specific chemical composition of the cloud from which the 16~Cyg binary system formed.

% C 16 Cyg - Sun =   -0.0720000 HD - Sun =    0.0650000
% O 16 Cyg - Sun =   -0.0710000 HD - Sun =    0.0700000
% Al 16 Cyg - Sun =    -0.115000 HD - Sun =     0.110000
% K 16 Cyg - Sun =    -0.118000 HD - Sun =     0.109000
% Zr 16 Cyg - Sun =   -0.0710000 HD - Sun =    0.0700000
% Ba 16 Cyg - Sun =    -0.115000 HD - Sun =     0.110000
% Nd 16 Cyg - Sun =    -0.118000 HD - Sun =     0.109000
%Na 16 Cyg - Sun =    -0.139000 HD - Sun =     0.129000
%S 16 Cyg - Sun =   -0.0950000 HD - Sun =    0.0860000
%Fe 16 Cyg - Sun =    -0.125000 HD - Sun =     0.106000
%Zn 16 Cyg - Sun =    -0.112000 HD - Sun =     0.106000

%\begin{equation}
%\Delta \mathrm{[X/H]}= \mathrm{log_{10}}\frac{f_\mathrm{X,photo}f_\mathrm{CZ}M_\mathrm{star}+f_\mathrm{X,acc}M_\mathrm{acc}}{f_\mathrm{X,photo}f_\mathrm{CZ}M_\mathrm{star}},
%\label{eq1}
%\end{equation}   

%\noindent
%where $f_\mathrm{X,photo}$ is the solar photospheric abundance of element X converted into mass fraction, $f_\mathrm{CZ}$ is the mass fraction of the convective envelope, and $f_\mathrm{X,acc}$ is the mass fraction of rocky element X in the accreted material. As in paper by \citet{2018ApJ...854..138O} we assume $f_\mathrm{CZ}$=0.02 \citep{2013ApJ...776...87S} and take the bulk Earth composition from \citet{2003TrGeo...2..547M}.
%{}2016ApJ...818...54M} {2018A&A...614A.138L}

\section{Conclusions}\label{sect:Conclusions}

This study deals with the two wide pairs: 16~Cyg~A and B, where the secondary component hosts a close-in giant planet, and HD~219542~A and B with no planet detected.
Atmospheric parameters of each star and the Sun were determined using the high-quality spectra and the \SME\ tools for automatic spectral analysis. Our method relies on a variety of spectroscopic indicators of effective temperature and surface gravity. A reliability of the method is confirmed by a perfect agreement of \te\ = 5778~K and \logg\ = 4.44 derived for the Sun with the canonical solar values.

By applying the synthetic spectrum method, we determined abundances of 29 chemical elements in each component of 16~Cyg, with the NLTE effects taken into account for 19 of them. Abundances of 23 chemical elements were derived for the HD~219542 binary components, with the NLTE abundances for 17 of them. The relative-to-solar abundances [X/H] were calculated by using a line-by-line differential approach. The solar abundances obtained from spectrum of the solar light reflected from the asteroid Vesta appear to be fairly consistent with the most recent solar photosphere \citep{2021A&A...653A.141A} and meteoritic \citep{2021SSRv..217...44L} abundances.

A careful analysis of the abundance differences between the components [A-B] and the  relative-to-solar abundances [X/H] for each binary system leads us to the following conclusions.

\begin{itemize}
	\item Both binary systems do not reveal significant abundance discrepancies between their A and B components, with the average abundance differences [A-B] = 0.019$\pm$0.012 and $-0.014\pm$0.019 for 16~Cyg and HD~219542, respectively.
	This provides an evidence for a negligible effect of the giant planet formation on atmospheric abundances of the host star. Although there is a hint of decreasing [A-B] for the refractory elements with \CT\ $>$ 1400~K in 16~Cyg.
	\item For both components of HD~219542, the trends of [X/H] with the dust condensation temperature are quasi-linear and similar to that by \citet{2009ApJ...704L..66M} for the solar analogues without detected planets.
	\item 16~Cyg reveals a different behaviour of [X/H]. Both components show a meaningful step-like change in [X/H] at \CT\ $\approx$ 750~K, then a decline with increasing \CT\ to about 1300~K, and splitting into two trends for the higher \CT. Such a similarity of the abundance trends for both components indicates a specific chemical composition of the cloud from which the 16~Cyg binary system formed.
\item The differences in the Li abundance between the components, 0.71~dex for 16~Cyg and 1.02~dex for
	HD~219542, can be explained by the differences in their effective temperatures.
\item From our stellar age estimates based on analysis of the isochrones, HD~219542 ($t$ = 3~Gyr) is younger than 16~Cyg ($t$ = 6~Gyr), and this explains its higher Li abundance.
\end{itemize}

\section*{Acknowledgments}
The authors acknowledge the support of Ministry of Science and Higher Education of the Russian Federation under the grant 075-15-2020-780 (N13.1902.21.0039). 
This research used the facilities of the Canadian Astronomy Data Centre operated by the National Research Council of Canada with the support of the Canadian Space Agency.
Based in part on data collected at Subaru Telescope and obtained from the SMOKA, which is operated by the Astronomy Data Center, National Astronomical Observatory of Japan.

\section*{Data Availability Statements}
The data underlying this article will be shared on reasonable request to the corresponding author.

\bibliographystyle{mnras}
\interlinepenalty=10000
\bibliography{paper}
\label{lastpage}
%\end{document}
%\Online
%--------------------------------------------------------------------
%--------------------------------------------------------------------
%\input{table_lines_1.tex}
%--------------------------------------------------------------------
%--------------------------------------------------------------------
\end{document}